\documentclass{llncs}
\usepackage{times}
\usepackage{amsfonts}
\usepackage{url}
\author{Yongge Wang}

\begin{document}

\title{Efficient Identity-Based and Authenticated Key Agreement Protocol}
\author{Yongge Wang}
\institute{UNC Charlotte (USA), {\tt yonwang@uncc.edu}}
\author{Yongge Wang}
\institute{UNC Charlotte (USA), {\tt yonwang@uncc.edu}}

\newcommand{\field}{\mathcal{P}}
\newcommand{\group}{\mathcal{G}}
\newcommand{\n}{\mathcal{\rho}}
\newcommand{\pr}{{\rm Pr}}
\newcommand{\Simu}{\mathcal{S}}
\newcommand{\A}{\mathcal{A}}
\newcommand{\U}{\mathcal{U}}
\newcommand{\C}{\mathcal{C}}
\newcommand{\D}{\mathcal{D}}
\newcommand{\G}{\mathcal{G}}
\newcommand{\X}{\mathcal{X}}
\newcommand{\Y}{\mathcal{Y}}
\newcommand{\R}{\mathcal{R}}
\newcommand{\Z}{\mathbb{Z}}
\newcommand{\ID}{\mathrm{ID}}
\newcommand{\GA}{\mathcal{A}^G}
\newcommand{\hash}{\mathcal{H}}
\newcommand{\oracle}{\mathcal{O}}
\newcommand{\pairing}{\hat{e}}

\maketitle
\begin{abstract}
Several identity based and implicitly authenticated key agreement 
protocols have been proposed in recent years and none of them 
has achieved all required security properties. 
In this paper, we propose an efficient identity-based and authenticated
key agreement protocol IDAK using Weil/Tate pairing. The security
of IDAK is proved in Bellare-Rogaway model. Several required 
properties for key agreement protocols are not implied by
the Bellare-Rogaway model. We proved these properties for IDAK 
separately. 
\end{abstract}

\section{Introduction}
\label{introduction}
Key establishment protocols are one of the most important 
cryptographic primitives that have been used in our society.
The first unauthenticated key agreement protocol based 
on asymmetric cryptographic techniques were proposed by 
Diffie and Hellman \cite{dh}. Since this seminal result, 
many authenticated key agreement protocols have been proposed 
and the security properties of key agreement protocols have been 
extensively studied. In order to implement these authenticated 
key agreement protocols, one needs to get the corresponding party's 
authenticated public key. For example, in order for Alice and Bob
to execute the NIST recommended MQV key agreement 
protocol \cite{mqv,fips},
Alice needs to get an authenticated 
public key $g^b$ for Bob  
and Bob needs to get an authenticated public key $g^a$ for Alice first,
where $a$ and $b$ are Alice and Bob's private keys respectively.  
One potential approach for implementing these schemes is to deploy 
a public key infrastructure (PKI) system, which has proven to 
be difficult. Thus it is preferred to design easy to deploy 
authenticated key agreement systems. Identity based key agreement 
system is such an example. 

In 1984, Shamir \cite{shamirid} proposed identity based cryptosystems
where user's identities (such as email address, phone numbers, 
office locations, etc.) could be used as the public keys.
Several identity based key agreement protocols (see, e.g., 
\cite{chen,gp90,mccullagh,ok86,sok00,scott,shim,smart,to91}) 
have been proposed since then.
Most of them are not practical or do not have all required security 
properties. Joux \cite{joux} proposed a one-round tripartite 
non-identity based key agreement protocol using Weil pairing. 
Then feasible identity based encryption schemes based 
on Weil or Tate paring were introduced
by Sakai, Ohgishi, and Kasahara \cite{sok00} and later by Boneh
and Franklin \cite{ibe} independently.

Based on Weil and Tate pairing techniques,
Smart \cite{smart}, Chen-Kudla \cite{chen},
Scott \cite{scott}, Shim \cite{shim}, and 
McCullagh-Barreto \cite{mccullagh} designed identity based
and authenticated key agreement protocols. 
Chen-Kudla \cite{chen} showed that Smart's protocol
is not secure in several aspects.  Cheng et al. \cite{cheng} pointed out 
that  Chen-Kudla's protocol is not secure againt unknown key share
attacks. Scott's protocol is not secure against man in the middle
attacks. Sun and Hsieh \cite{sun} showed that Shim's 
protocol is insecure against key compromise impersonation attacks
or man in the middle attacks. Choo \cite{kwang} showed that 
McCullagh and Barreto's protocol is  insecure against key revealing attacks.
McCullagh and Barreto \cite{mccullaghr} revised their protocol.
But the revised protocol does not achieve weak perfect forward
secrecy property.
In this paper, we propose
an efficient identity based and authenticated key agreement protocol 
achieving all security properties that an authenticated key agreement 
protocol should have. 

The advantage of identity based key agreement is that non-PKI system
is required. The only prerequisite for executing identity based key 
agreement protocols is the deployment of authenticated system-wide 
parameters. Thus, it is easy to implement these protocols 
in relatively closed environments such as government organizations 
and commercial entities.


The remainder of this paper is organized as follows. In \S\ref{bilinear}
we briefly describe bilinear maps, bilinear Diffie-Hellman problem,
and its variants. In \S\ref{idakprotocol}, we describe our identity based
and authenticated key agreement protocol IDAK. 
\S\ref{securitymodel} describes a security model for identity based key 
agreement.
In section \S\ref{securityproof}, we prove the security of 
IDAK key agreement protocol.  In sections \S\ref{pfsidak} and 
\S\ref{kcridak}, we discuss key  compromise impersonation resilience
and perfect forward secrecy properties of IDAK key agreement 
protocol. 

\section{Bilinear maps and the bilinear Diffie-Hellman assumptions}
\label{bilinear}
In the following, we briefly describe the bilinear maps
and bilinear map groups. The details could be found 
in Joux \cite{joux} and Boneh and Franklin \cite{ibe}.
\begin{enumerate}
\item $G$ and $G_1$ are two (multiplicative) cyclic groups of
prime order $q$.
\item $g$ is a generator of $G$.
\item $\pairing:G\times G\rightarrow G_1$ is a bilinear map.
\end{enumerate}
A bilinear map is a map $\pairing:G\times G\rightarrow G_1$ 
with the following properties:
\begin{enumerate}
\item bilinear: for all $g_1,g_2\in G$, and $x,y\in Z$, we have 
$\pairing(g_1^x,g_2^y)=\pairing(g_1,g_2)^{xy}$.
\item non-degenerate: $\pairing(g,g)\not=1$.
\end{enumerate}
We say that $G$ is a bilinear group if the group action in $G$ can be 
computed efficiently and there exists a group $G_1$ and an efficiently
computable bilinear map $\pairing:G\times G\rightarrow G_1$ as above.
Concrete examples of bilinear groups are given in \cite{joux,ibe}.
For convenience, throughout the paper, we view both $G$ and $G_1$ 
as multiplicative groups though the concrete implementation of $G$ 
could be additive elliptic curve groups.

Throughout the paper {\em efficient} means
probabilistic polynomial-time, {\em negligible} refers to a function
$\varepsilon_k$ which is smaller than $1/k^{c}$ for 
all ${c}>0$ and sufficiently large $k$, and
{\em overwhelming} refers to a function $1-\varepsilon_k$
for some negligible $\varepsilon_k$. Consequently, a function
$\delta_k$ is {\em non-negligible} if there exists a constant $c$ and there 
are infinitely many $k$ such that $\delta_k >1/k^c$.
We first formally define the notion of a bilinear group family and 
computational indistinguishable distributions (some of 
our terminologies are adapted from Boneh \cite{boneh}).

\noindent
{\bf Bilinear group families} A {\em bilinear group family} 
$\group$ is a set $\group=\{G_\n\}$ of bilinear groups 
$G_\n=\langle G, G_1, \pairing\rangle$ 
where $\n$ ranges over an infinite index set, $G$ and $G_1$ are two 
groups of prime order $q_\n$, and $\pairing: G\times G\rightarrow G_1$ 
is a bilinear map. We denote by $|\n|$ 
the length of the binary representation
of $\n$. We assume that group and bilinear operations 
in $G_\n=\langle G, G_1, \pairing\rangle$ are efficient in $|\n|$.
Unless specified otherwise, we will abuse
our notations by using $q$ as the group order instead of $q_\n$ 
in the remaining part of this paper.

\noindent
{\bf Instance generator} An {\em Instance Generator}, $\mathcal{IG}$,
for a bilinear group family $\group$ is a randomized algorithm that given
an integer $k$ (in unary, that is, $1^k$), runs in polynomial-time in $k$ 
and outputs some random index $\n$ for 
$G_\n=\langle G, G_1, \pairing\rangle$, and a generator $g$ of $G$, 
where $G$ and $G_1$ are groups of prime order $q$. 
Note that for each $k$, the Instance Generator induces a distribution
on the set of indices $\n$. 

The following Bilinear Diffie-Hellman Assumption (BDH) has been used by
Boneh and Franklin \cite{ibe} to show security of their
identity-based encryption scheme.

\noindent
{\bf Bilinear Diffie-Hellman Problem}  
Let $\group=\{G_\n\}$ be a bilinear group family and 
$g$ be a generator for $G$, where $G_\n=\langle G, G_1, \pairing\rangle$. 
The BDH problem in
$\group$ is as follows: given $\langle g, g^x, g^y, g^z\rangle$ 
for some $x,y,z\in Z_q^*$, compute
$\pairing (g,g)^{xyz}\in G_1$. A CBDH algorithm $\C$ 
for $\group$ is a probabilistic polynomial-time 
algorithm that can compute the function 
$\mathrm{BDH}_g(g^x,g^y,g^z)=\pairing(g,g)^{xyz}$ 
in $G_\n$ with a non-negligible probability.
That is, for some fixed ${c}$ we have
\begin{equation}
\label{cbdhe}
\pr\left[\C(\n,g,g^x,g^y,g^z)=\pairing(g,g)^{xyz}\right]\ge
\frac{1}{k^{c}}
\end{equation}
where the probability is over the random choices of
$x,y,z$ in $Z_q^*$, the index $\n$,
the random choice of $g\in G$, and the random bits of  $\A$.

\noindent
{\bf CBDH Assumption}. 
The bilinear group family $\group=\{G_\n\}$ {\em satisfies}
the CBDH-Assumption if there is no CBDH algorithm for 
$\group$. A perfect-CBDH algorithm $\C$ 
for $\group$ is a probabilistic polynomial-time 
algorithm that can compute the function
$\mathrm{BDH}_g(g^x,g^y,g^z)=\pairing(g,g)^{xyz}$ 
in $G_\n$ with overwhelming probability. $\group$ {\em satisfies}
the perfect-CBDH-Assumption if there is no perfect-CBDH algorithm 
for $\group$.

\begin{theorem}
\label{perfectcbdh}
A bilinear group family $\group$ satisfies the CBDH-Assumption if and only 
if it satisfies the perfect-CBDH-Assumption.  
\end{theorem}

\noindent
{\bf Proof.} See Appendix.
\hfill$\Box$

Consider Joux's tripartite key agreement protocol \cite{joux}:
Alice, Bob, and Carol fix a bilinear group $\langle G, G_1, \pairing\rangle$.
They select $x,y,z\in_R Z_q^*$ and exchange $g^x$, $g^y$, and $g^z$. 
Their shared secret is $\pairing(g,g)^{xyz}$. To {\em totally
break} the protocol a passive eavesdropper, Eve, must
compute the BDH function: $\mathrm{BDH}_g(g^x,g^y,g^z)=\pairing(g,g)^{xyz}$.

CBDH-Assumption by itself is not sufficient to prove
that Joux's protocol is useful for practical
cryptographic purposes. Even though Eve may be unable 
to recover the entire secret, she may still be able 
to predict quite a few bits (less than $c\log k$ bits for some constant $c$;
Otherwise, CBDH assumption is violated) of information 
for $\pairing(g,g)^{xyz}$ with some confidence. 
If $\pairing(g,g)^{xyx}$ is to be the basis of a shared secret key, one must
bound the amount of information Eve is able to deduce about it, 
given $g^x$, $g^y$, and $g^z$. This is formally captured by the,
much stronger, Decisional Bilinear Diffie-Hellman assumption 
(DBDH-Assumption)
\begin{definition}
\label{dis}
Let $\{\X_\n\}$ and $\{\Y_\n\}$
be two ensembles of probability distributions, where
for each $\n$ both $\X_\n$ and $\Y_\n$
are defined over the same domain. We say
that the two ensembles are {\em computationally indistinguishable}
if for any probabilistic polynomial-time algorithm  $\D$,
and any ${c}>0$ we have 
$$\left|\pr\left[\D\left(\X_\n\right)
=1\right]-
\pr\left[\D\left(\Y_\n\right)=1\right]
\right|<\frac{1}{k^{c}}$$
for all sufficiently large $k$, where the probability
is taken over all $\X_\n$, $\Y_\n$, and internal 
coin tosses of $\D$.
\end{definition}
In the remainder of the paper,
we will say in short that the two distributions  $\X_\n$ and $\Y_\n$ are
computationally indistinguishable.

Let $\group=\{G_\n\}$ be a bilinear group family. We consider the following 
two ensembles of distributions:
\begin{itemize}
\item $\{\X_\n\}$
of random tuples $\langle \n, g, g^x,g^y,g^z,\pairing(g,g)^t \rangle$,
where $g$ is a random generator of $G$ 
($G_\n=\langle G, G_1, \pairing\rangle$) and $x,y,z,t\in_R Z_q$.
\item $\{\Y_\n\}$ of tuples
$\langle \n, g, g^x,g^y,g^z, \pairing(g,g)^{xyz}\rangle$, where 
$g$ is a random generator of $G$ and $x,y,z\in_R Z_q$.
\end{itemize}

An algorithm that solves the Bilinear Diffie-Hellman decision
problem is a polynomial time probabilistic algorithm 
that can effectively distinguish
these two distributions. That is, given a tuple coming
from one of the two distributions, it should output 0 or 1,
and there should be a non-negligible difference between
(a) the probability that it outputs a 1 given an input
from $\{\X_\n\}$, and (b) the probability that it outputs a 1 
given an input from $\{\Y_\n\}$. The bilinear group family $\group$
{\em satisfies the DBDH-Assumption} if 
the two distributions are computationally indistinguishable.

\noindent
{\bf Remark.} The DBDH-Assumption is implied by a slightly
weaker assumption: {\em perfect}-DBDH-Assumption.
A perfect-DBDH statistical test for $\group$ distinguishes
the inputs from the above $\{\X_\n\}$ and $\{\Y_\n\}$ with
overwhelming probability. The bilinear group family $\group$ 
{\em satisfies the perfect-DBDH-Assumption} 
if there is no such probabilistic polynomial-time statistical test.

\section{The scheme IDAK}
\label{idakprotocol}
In this section, we describe our identity-based and authenticated 
key agreement scheme IDAK. 
Let $k$ be the security parameter given to the setup algorithm 
and $\mathcal{IG}$ be a bilinear group parameter generator.
We present the scheme by describing the
three algorithms: {\bf Setup}, {\bf Extract}, and {\bf Exchange}. 

\vskip 3pt
\noindent
{\bf Setup}: For the input $k\in Z^{+}$, 
the algorithm proceeds as follows:
\begin{enumerate}
\item Run $\mathcal{IG}$ on $k$ to generate a bilinear group
$G_\n=\{G,G_1,\pairing\}$ and the prime order $q$ of the two groups 
$G$ and $G_1$. 
\item Pick a random master secret $\alpha\in  Z^*_q$.
\item Choose cryptographic hash functions $H:\{0,1\}^*\rightarrow G$
and $\pi:G\times G\rightarrow Z_q^*$. In the security analysis, 
we view $H$ and $\pi$ as random oracles. In practice, 
we take  $\pi$ as a random oracle (secure hash function) from 
$G\times G$ to $Z_{2^{\lceil\log q\rceil/2}}^*$ (see Appendix for details).
\end{enumerate}
The system parameter is $\langle q, g, G, G_1, \pairing, 
H,\pi\rangle$ and the master secret key is $\alpha$.

\vskip 3pt
\noindent
{\bf Extract}: For a given identification 
string $\ID\in \{0,1\}^*$, the algorithm 
computes a generator $g_{\ID}=H(\ID)\in G$, and sets the private key 
$d_{\ID}=g_{\ID}^\alpha$ where $\alpha$ is the master secret key.

\vskip 3pt
\noindent
{\bf Exchange}: For two participants Alice and Bob whose identification
strings are $\ID_A$ and $\ID_B$ respectively, 
the algorithm proceeds as follows.
\begin{enumerate}
\item Alice selects $x\in_R Z_q^*$, 
computes $R_A=g_{\ID_A}^{x}$, and sends it to Bob.
\item  Bob selects $y\in_R Z_q^*$, 
computes $R_B=g_{\ID_B}^{y}$, and sends it to Alice.
\item Alice computes
$s_A=\pi(R_A,R_B)$, $s_B=\pi(R_B,R_A)$, and 
 the shared secret $sk_{AB}$ as
$$\pairing(g_{\ID_A}, g_{\ID_B})^{(x+s_A)(y+s_B)\alpha}=
\pairing\left(d_{\ID_A}^{(x+s_A)}, g_{\ID_B}^{s_B}\cdot R_B\right).$$
\item Bob computes $s_A=\pi(R_A,R_B)$, $s_B=\pi(R_B,R_A)$, and  
the shared secret $sk_{BA}$ as 
$$\pairing(g_{\ID_A}, g_{\ID_B})^{(x+s_A)(y+s_B)\alpha}=
\pairing\left(g_{\ID_A}^{s_A}\cdot 
R_A, d_{\ID_B}^{(y+s_B)}\right).$$
\end{enumerate}

In the next section, we will show that IDAK protocol
is secure in Bellare and Rogaway \cite{br1} model with
random oracle plus DBDH-Assumption. We conclude this section with a 
theorem which says that the shared secret established by the IDAK key 
agreement protocol is computationally indistinguishable from a random value.

\begin{theorem}
\label{passiverandom}
Let $\group=\{G_\n\}$ be a bilinear group family,
$G_\n=\langle G, G_1, \pairing\rangle$, and $g_1,g_2$ be random generators
of $G$. Assume that DBDH-Assumption holds for $\group$.
Then the distributions $\langle g_1,g_2$, $g_1^x,g_2^y$, 
$\pairing(g_1,g_2)^{(x+\pi(g_1^x,g_2^y))(y+\pi(g_2^y,g_1^x))\alpha}\rangle$ 
and $\langle g_1,g_2,g_1^x,g_2^y, \pairing(g_1,g_2)^{z}\rangle$ 
are computationally indistinguishable, where $\alpha,x,y,z$ are selected
from $Z^*_q$ uniformly. 
\end{theorem}

Before we give a proof for  Theorem \ref{passiverandom}, we first 
prove two lemmas that will be used in the proof of the Theorem.

\begin{lemma}
\label{firstlemma}
(Naor and Reingold \cite{naor}) 
Let $\group=\{G_\n\}$ be a bilinear group family,
$G_\n=\langle G, G_1, \pairing\rangle$, $m$ be a constant, 
$g$ be a random generator of $G$, and $\hat{g}=\pairing(g,g)$.
Assume that the DBDH-Assumption holds for $G_\n$.
Then the two distributions
$\langle \R, (\hat{g}^{x_iy_jz_l}:i, j,l\le m)\rangle$
and $\langle \R, (\hat{g}^{u_{ijl}}: i,j,l\le m)\rangle$
are computationally indistinguishable.
Here $\R$ denotes the tuple $(g, (g^{x_i}, g^{y_j}, g^{z_l}: i,j,l\le m))$ 
and $x_i,y_j, z_{l}, u_{ijl}\in_R Z_q$.
\end{lemma}
\noindent
{\bf Proof.} Using a random reduction,
Naor and Reingold \cite[Lemma 4.4]{naor} (see also Shoup \cite[\S5.3.2]{shoup}
showed that the two distributions
$\langle \R, (g^{x_iy_j}:i, j\le m)\rangle$
and $\langle \R, (g^{u_{ij}}: i, j\le m)\rangle$
are computationally indistinguishable. 
The proof can be directly modified to obtain a proof for this Lemma.
The details are omitted.
\hfill$\Box$

\begin{lemma}
\label{secondlemma}
Let $\group=\{G_\n\}$ be a bilinear group family,
$G_\n=\langle G, G_1, \pairing\rangle$, $g$ be a random generator of $G$,
$\hat{g}=\pairing(g,g)$, and $f_1$ and $f_2$ be two 
polynomial-time computable functions. If the two distributions
$\X_1= \langle \R, \hat{g}^{f_1(\mathbf{x})}, 
\hat{g}^{f_2(\mathbf{x})}\rangle$
and 
$\Y_1=\langle \R, \hat{g}^{z_1}, \hat{g}^{z_2}\rangle$
are computationally indistinguishable, then the two distributions 
$\X_2=\langle \R_1, 
\hat{g}^{f_1(\mathbf{x})+f_2(\mathbf{x})}\rangle$
and 
$\Y_2=\langle \R_2, \hat{g}^{z}\rangle$
are computationally indistinguishable, where 
$\R=\left(g, \left(g^{x_i}: 1\le i\le m\right)\right)$, $\mathbf{x}=
(x_1,\ldots, x_m)$, and 
$x_i, z_1,z_2,z\in_R Z_q$.
\end{lemma}

\noindent
{\bf Proof.} See Appendix.
\hfill$\Box$

\vskip 5pt
\noindent
{\bf Proof of Theorem \ref{passiverandom}} Let $\hat{g}=\pairing(g,g).
$By Lemma \ref{firstlemma}, the two distributions
$$\begin{array}{l}
\X=\langle g, g^\alpha,g^x,g^y, \hat{g}^{\alpha xy}, 
\hat{g}^{\alpha x\pi(g^y,g^x)},
   \hat{g}^{\alpha y\pi(g^x,g^y)},
   \hat{g}^{\alpha \pi(g^x,g^y)\pi(g^y,g^x)}\rangle\quad\mbox{and}\\
\Y=\langle g, g^\alpha,g^x,g^y, \hat{g}^{z_1'}, \hat{g}^{z_2'\pi(,g^y,g^x)},
   \hat{g}^{z_3'\pi(g^x,g^y)},\hat{g}^{z_4'\pi(g^x,g^y)\pi(g^y,g^x)}\rangle
\end{array}$$
are computationally indistinguishable assuming that
DBDH-Assumption holds for $\group$, where $g$ is 
a random generator of $G_\n$ and $\alpha,x$, $y$, 
$z_1'$, $z_2'$, $z_3',z_4'\in_R Z_q$. 
Since $\pi$ is a fixed function from $G$ to $Z_q^*$
and $q$ is a prime,  it is straightforward to verify 
that for any $\alpha,x,y\in Z_q$, 
$\hat{g}^{z_2'\pi(g^y,g^x)}$, $\hat{g}^{z_3'\pi(g^x,g^y)}$, 
and $\hat{g}^{z_4'\pi(g^x,g^y)\pi(g^y,g^x)}$ 
are uniformly (and independently of
each other) distributed over $G_1$. It follows that the distribution
$$\mathcal{Z}=
\langle g, g^\alpha,g^x,g^y, \hat{g}^{z_1}, \hat{g}^{z_2},
   \hat{g}^{z_3},\hat{g}^{z_4})\rangle$$
is computationally indistinguishable from the 
distribution $\Y$, where $z_1,z_2,z_3,z_4\in_R Z_q$.
Thus $\X$ and $\mathcal{Z}$ are computationally indistinguishable.
The Theorem now follows from Lemma \ref{secondlemma}.
\hfill$\Box$

\section{The security model}
\label{securitymodel}
Our security model is based on Bellare and Rogaway \cite{br1} 
security models for key agreement protocols with several modifications.
In our model, we assume that we have at most $m\le \mbox{poly}(k)$ 
protocol participants (principals): $\ID_1, \ldots, \ID_m$, 
where $k$ is the security parameter. 
The protocol determines how principals
behave in response to input signals from their environment. Each principal
may execute the protocol multiple times with the same or different
partners. This is modelled by allowing each principal to have different
instances that execute the protocol. An oracle $\Pi_{i,j}^s$ models the 
behavior of the principal $\ID_i$ carrying 
out a protocol session in the 
belief that it is communicating with the principal $\ID_j$ for 
the $s$th time. One given instance is used only for one time. 
Each $\Pi_{i,j}^s$ maintains a variable {\em view} (or {\em transcript}) 
consisting of the protocol run transcripts so far.

The adversary is modelled by a probabilistic polynomial time
Turing machine that is assumed to have complete control over 
all communication links in the network and to interact with 
the principals via oracle accesses to $\Pi_{i,j}^s$. The adversary
is allowed to execute any of the following queries:
\begin{itemize}
\item ${\bf Extract}(\ID)$. This allows the adversary to get the
long term private key for a new principal whose identity 
string is $\ID$.
\item ${\bf Send}(\Pi_{i,j}^s, X)$. This sends message $X$ to the oracle
$\Pi_{i,j}^s$. The output of $\Pi_{i,j}^s$ is given to the adversary.
The adversary can ask the principal $\ID_i$ to initiate a session with 
$\ID_j$ by a query ${\bf Send}(\Pi_{i,j}^s, \lambda)$ where $\lambda$
is the empty string.
\item ${\bf Reveal}(\Pi_{i,j}^s)$. This asks the oracle to reveal whatever
session key it currently holds.
\item ${\bf Corrupt}(i)$. This asks $\ID_i$ to reveal
the long term private key $d_{\ID_i}$. 
\end{itemize}
The difference between the queries {\bf Extract} and {\bf Corrupt}
is that the adversary can use {\bf Extract} to get the private key 
for an identity string of her choice while {\bf Corrupt} can only be used
to get the private key of existing principals.

Let $\Pi_{ij}^s$ be an initiator oracle (that is, it has received a 
$\lambda$ message at the beginning) and $\Pi_{ji}^{s'}$ be a responder
oracle. If every message that $\Pi_{ij}^s$ sends out is subsequently
delivered to $\Pi_{ji}^{s'}$, with the response to this message
being returned to $\Pi_{ij}^s$ as the next message on its transcript, then
we say the oracle $\Pi_{ji}^{s'}$ matches $\Pi_{ij}^s$. Similarly, if every
message that $\Pi_{ji}^{s'}$ receives was previously generated by
$\Pi_{ij}^s$, and each message that $\Pi_{ji}^{s'}$ sends out is subsequently
delivered to $\Pi_{ij}^s$, with the response to this message
being returned to $\Pi_{ji}^{s'}$ as the next message on its transcript, then
we say the oracle $\Pi_{ij}^s$ matches $\Pi_{ji}^{s'}$.
The details for an exact definition of matching oracles 
could be found in \cite{br}. 

For the definition of matching oracles, the reader should be 
aware the following scenarios: Even though the oracle $\Pi_{ij}^s$ 
thinks that its matching oracle is $\Pi_{ji}^{s'}$, the real matching
oracle for $\Pi_{ij}^s$ could be $\Pi_{ji}^{t'}$. For example,
if $\Pi_{ij}^s$  sends a message $X$ to $\Pi_{ji}^{s'}$ and $\Pi_{ji}^{s'}$
replies with $Y$. The adversary decides not to forward the message 
$Y$ to $\Pi_{ij}^s$. Instead, the adversary sends the message $X$
to initiate another oracle $\Pi_{ji}^{t'}$ and $\ID_i$ does not know 
the existence of this new oracle $\Pi_{ji}^{t'}$. The oracle $\Pi_{ji}^{t'}$
replies with $Y'$ and the adversary forwards this $Y'$ to $\Pi_{ij}^s$ 
as the responding message for $X$. In this case, the transcript
of $\Pi_{ij}^s$  matches the transcript of $\Pi_{ji}^{t'}$. Thus we 
consider $\Pi_{ij}^s$ and $\Pi_{ji}^{t'}$ as matching oracles. 
In another word, the matching oracles are mainly based the message
transcripts.

In order to define the notion of a secure session key exchange,
the adversary is given an additional experiment.
That is, in addition to the above regular queries,
the adversary can choose, at any time during its run, a 
${\bf Test}(\Pi_{i,j}^s)$ query to a completed oracle $\Pi_{i,j}^s$
with the following properties:
\begin{itemize}
\item The adversary has never issued,  at any time during its run, the 
query ${\bf Extract}(\ID_i)$ or ${\bf Extract}(\ID_j)$. 
\item The adversary has never issued,  at any time during its run, the 
query ${\bf Corrupt}(i)$ or ${\bf Corrupt}(j)$. 
\item The adversary has never issued,  at any time during its run, the 
query ${\bf Reveal}(\Pi_{i,j}^s)$.
\item  The adversary has never issued,  at any time during its run, the 
query ${\bf Reveal}(\Pi_{j,i}^{s'})$ if the matching oracle
$\Pi_{j,i}^{s'}$ for $\Pi_{i,j}^s$ exists (note that such an oracle
may not exist if the adversary is impersonating the $\ID_j$
to the oracle $\Pi_{i,j}^s$). The value of $s$ may be different from the 
value of $s'$ since the adversary may run fake sessions to 
impersonate any principals without victims' knowledge.
\end{itemize}
Let $sk^s_{i,j}$ be the value of the session key held by the 
oracle $\Pi_{i,j}^s$ that has been established between 
$\ID_i$ and $\ID_j$.
The oracle $\Pi_{i,j}^s$ tosses a coin $b\leftarrow_R\{0,1\}$.
If $b=1$, the adversary is given $sk^s_{i,j}$. Otherwise, 
the adversary is given a value $r$ randomly chosen from the probability
distribution of keys generated by the protocol. In the end,
the attacker outputs a bit $b'$. The advantage that the adversary
has for the above guess is defined as 
$$\mathrm{Adv}^{\A}(k)=\left|\Pr[b=b']-\frac{1}{2}\right|.$$
Now we are ready to give the exact definition for a secure 
key agreement protocol.

\begin{definition}
\label{keysecuredef}
A key agreement protocol $\Pi$ is BR-secure if the following conditions
are satisfied for any adversary:
\begin{enumerate}
\item If two uncorrupted oracles $\Pi_{ij}^s$ and $\Pi_{ji}^{s'}$
have matching conversations (e.g., the adversary is passive)
and both of them are complete according
to the protocol $\Pi$, then both oracles will always 
accept and hold the same session key which is uniformly distributed
over the key space. 
\item $\mathrm{Adv}^{\A}(k)$ is negligible.
\end{enumerate}
\end{definition}

In the following, we briefly discuss the attributes that a BR-secure
key agreement protocol achieves.
\begin{itemize}
\item {\bf Known session keys}. The adversary may use 
{\bf Reveal}$(\Pi_{i,j}^{s'})$ query before or after the query 
{\bf Test}$(\Pi_{i,j}^s)$. Thus in a secure key agreement model,
the adversary learns zero information about a fresh
key for session $s$ even if she has learnt keys for other sessions 
$s'$.
\item {\bf Impersonation attack}. If the adversary impersonates 
$\ID_j$ to $\ID_i$, then she still learns 
zero information about the session key that the oracle 
$\Pi_{ij}^s$ holds for this impersonated $\ID_j$ since 
there is no matching oracle for $\Pi_{ij}^s$ in this scenario. 
Thus $\A$ can use {\bf Test} query to test this session key that
$\Pi_{ij}^s$ holds. 
\item {\bf Unknown key share}. If $\ID_i$ establishes 
a session key with $\ID_l$ though he believes that he is 
talking to $\ID_j$, then there is an oracle $\Pi_{ij}^s$
that holds this session key $sk_{ij}$. At the same time, there is an
oracle $\Pi_{li'}^{s'}$ that holds this session key $sk_{ij}$, for some
$i'$ (normally $i'=i$). During an unknown key share attack, 
the user $\ID_j$ may not know this session key. 
Since $\Pi_{ij}^s$ and $\Pi_{li'}^{s'}$ are not matching oracles, 
the adversary can make the query ${\bf Reveal}(\Pi_{li'}^{s'})$
to learn this session key before the query ${\bf Test}(\Pi_{ij}^s)$.
Thus the adversary will succeed for this {\bf Test} query challenge
if the unknown key share attack is possible.
\end{itemize}
However, the following important security properties that a secure key 
agreement scheme should have are not implied from the original
BR-security model.
\begin{itemize}
\item {\bf Perfect forward secrecy}. This property requires that
previously agreed session keys should remain secret, even if both
parties' long-term private key materials are compromised.
Bellare-Rogaway model does not capture this property. 
Canetti and Krawczyk's model \cite{ck} use the session-key expiration
primitive to capture this property. Similar modification 
to Bellare-Rogaway model are required to capture this property also.
We will give a separate proof that the IDAK key agreement protocol achieves
weak perfect forward secrecy. Note that as pointed
out in \cite{hmqv}, no two-message key-exchange protocol 
authenticated with public keys and with no secure shared state 
can achieve perfect forward secrecy.
\item {\bf Key compromise impersonation resilience}. If the entity $A$'s
long term private key is compromised, then the adversary could 
impersonate $A$ to others, but it should not be able to impersonate 
others to $A$. Similar to wPFS property, Bellare-Rogaway 
model does not capture this property. We will give a 
separate proof that the IDAK key agreement protocol has this property.
\end{itemize}

\section{The security of IDAK}
\label{securityproof}
Before we present the security proof for the IDAK key agreement
protocol, we first prove some preliminary results that will be used in the
security proof. 

\begin{lemma}
\label{feedbackBDH}
Let $\group=\{G_\n\}$ be a bilinear group family,
$G_\n=\langle G, G_1, \pairing\rangle$, $g$ be a random 
generator of $G$, and $\pi:G\times G\rightarrow Z_q$ be a random
oracle. Assume DBDH-Assumption holds for $\group$ and let $\X$ and $\Y$
be two distributions defined as 
$$\begin{array}{ll}
&\X=\langle \R,g^{\beta x_0}, g^{\gamma y_0},
\pairing(g,g)^{(x_0+\pi(g^{\beta x_0},g^{\gamma y_0}))
(y_0+\pi(g^{\gamma y_0},g^{\beta x_0}))\alpha\beta\gamma},
           \pairing(g,g)^{\alpha\beta\gamma}\rangle \\ 
\mbox{and}\quad\quad&
\Y=\langle \R, g^{\beta x_0}, g^{\gamma y_0},
\pairing(g,g)^{(x_0+\pi(g^{\beta x_0},g^{\gamma y_0}))
(y_0+\pi(g^{\gamma y_0},g^{\beta x_0}))t},\pairing(g,g)^{t}\rangle
\end{array}$$ 
Then we have
\begin{enumerate}
\item The two distributions $\X$ and $\Y$ are computationally 
indistinguishable if $\R$ is defined as
$$\R=\left(g, g^\alpha, g^\beta, g^\gamma, g^{x}, g^r, g_\A,
\pairing\left(g^{x+\beta\pi({g^{x},g_\A})},
g_\A\cdot g^{r\pi(g_\A,g^{x})}\right)^{\alpha}\right),$$
$\alpha, \beta, \gamma, x, t, x_0$ are chosen from $Z_q^*$ uniformly,
    $g^r=g^\gamma$ or $r$ is either chosen from $Z_q^*$ uniformly,
$g_\A$ and $g^{\gamma y_0}$ are chosen from $G$ 
within polynomial time according to a fixed distribution given the view 
$({g^{x},g^r,g^\alpha, g^\beta, g^\gamma,g^{\beta x_0}})$
without violating DBDH-Assumption.
\item For any constant $m\le \mbox{poly}(k)$, 
the two distributions $\X$ and $\Y$ are computationally 
indistinguishable if $\R$ is defined as:
$$(g, g^\alpha, g^\beta, g^\gamma, (g^{x_i}, g^{r_j}, g_{\A,l})_{i,j,l\le m}, 
(\pairing(g^{x_i+\beta\pi(g^{x_i},g_{\A,l})},
g_{\A,l}\cdot g^{r_j\pi(g_{\A,l},g^{x_i})})^\alpha: 
i,j,l\le m))$$
where $\alpha, \beta, \gamma, x_i$ are uniformly chosen from 
$Z_q^*$, $r_j$ are either chosen from $Z_q^*$ uniformly or $g^{r_j}=g^\gamma$,
and $g_{\A,l}$ is chosen 
within polynomial time according to a fixed distribution given the view 
$(g^{x_i}$, $g^{r_j}$, $g^\alpha$, $g^\beta,g^\gamma, g^{\beta x_0}: 
i,j,l\le m)$ without violating DBDH-Assumption.
\item For any constant $m\le \mbox{poly}(k)$, 
the two distributions $\X$ and $\Y$ are computationally 
indistinguishable if $\R=({\R_1,\R_2})$, where 
$\R_1$ is defined as the $\R$ in the item 2, and 
$\R_2$ is defined as:
$$((g_{\A, i}, g^{r_j}, g_{\A,l})_{i,j,l\le m}, 
(\pairing(g_{\A, i}\cdot g^{\beta\pi(g_{\A, i},g_{\A,l})},
g_{\A,l}\cdot g^{r_j\pi(g_{\A,l},g_{\A, i})})^\alpha: 
i,j,l\le m))$$
where $r_j$ are either chosen from $Z_q^*$ uniformly or $g^{r_j}=g^\gamma$,
$g_{\A, i}$ and $g_{\A,l}$ are chosen 
within polynomial time according to a fixed distribution given the view 
$(g^{x_i}$, $g^{r_j}$, $g^\alpha$, $g^\beta,g^\gamma, g^{\beta x_0}, 
g^{\gamma y_0}: i,j,l\le m)$ without violating DBDH-Assumption
and with the condition that 
``$g_{\A, i}\not=g^{\beta x_0}$ or $g_{\A,l}\not=g^{\gamma y_0}$''.
Note that $g_{\A, i}$ and $g_{\A,l}$ could have different
distributions.
\end{enumerate}
\end{lemma}

\noindent
{\bf Proof.} See Appendix.\hfill$\Box$

\begin{theorem}
\label{securityprooftheorem}
Suppose that the functions $H$ and $\pi$ are random oracles and
the bilinear group family $\group$ satisfies DBDH-Assumption. Then
the IDAK scheme is a BR-secure key agreement protocol.
\end{theorem}

\noindent
{\bf Proof.} See Appendix.\hfill$\Box$

\section{Weak Perfect forward secrecy}
\label{pfsidak}
In this section, we show that the protocol IDAK 
achieves weak perfect forward secrecy property. 
Perfect forward secrecy property requires that even if Alice and Bob 
lose their private keys $d_{ID_A}=g_{ID_A}^\alpha$ and 
$d_{ID_B}=g_{\ID_B}^\alpha$, the session keys  established by Alice 
and Bob in the previous sessions are still secure. 
Krawczyk \cite{hmqv} pointed out that 
no two-message key-exchange protocol authenticated with 
public keys and with no secure shared state 
can achieve perfect forward secrecy. 
Weak perfect forward secrecy (wPFS) property for key agreement protocols
sates as follows \cite{hmqv}: any session key 
established by uncorrupted parties without active intervention by the
adversary is guaranteed to remain secure even if the parties to
the exchange are corrupted after the session key was erased
from the parties memory (for a formal definition, 
the reader is referred to \cite{hmqv}).

In the following, we show the IDAK achieves wPFS property.
Using the similar primitive of ``session-key expiration'' 
as in Canetti and Krawczyk's model \cite{ck},
we can revise Bellare-Rogaway model so that wPFS property
is provable also. In Bellare-Rogaway model,
the ${\bf Test}(\Pi_{i,j}^s)$ query is allowed only if the four
properties in Section \ref{securitymodel} are satisfied.
We can replace the property 
``the adversary has never issued,  at any time during its run, the 
query ${\bf Corrupt}(i)$ or ${\bf Corrupt}(j)$'' with the
property ``the adversary has never issued,  before the session
$\Pi_{i,j}^s$ is complete, the 
query ${\bf Corrupt}(i)$ or ${\bf Corrupt}(j)$''.
We call this model the wpfsBR model. In the 
final version of this paper, we will show that the protocol IDAK 
is secure in the wpfsBR model. Thus IDAK achieves wPFS property.
In the following, we present the essential technique used in the proof.
It is essentially sufficient to show that 
the two distributions $\left(\R, \pairing(g_{\ID_A},g_{\ID_B})^{z}\right)$ 
and $\left(\R, \pairing(g_{\ID_A},
g_{\ID_B})^{(x+\pi(g_{\ID_A}^x, g_{\ID_B}^y))(y+\pi(g_{\ID_B}^y,
g_{\ID_A}^x))\alpha}\right)$ 
are computationally indistinguishable for 
$\R=(g_{\ID_A}^\alpha, g_{\ID_B}^\alpha,g_{\ID_A}^x,g_{\ID_B}^y)$
and uniform at random chosen $g_{\ID_A}$, $g_{\ID_B}$, $x,y,z,\alpha$.
Consequently, it is sufficient to prove the following theorem.

\begin{theorem}
\label{pfsthm}
Let $\group=\{G_\n\}$ be a bilinear group family,
$G_\n=\langle G, G_1, \pairing\rangle$. 
Assume that DBDH-Assumption holds for $\group$.
Then the two distributions
$$\begin{array}{ll}
&\X=\left(g_1, g_2,g_1^\alpha, g_2^\alpha,g_1^x,
g_2^y, \pairing(g_1, g_2)^{xy\alpha}\right)\\
\mbox{and}\quad\quad &
\Y=\left(g_1, g_2,g_1^\alpha, g_2^\alpha,g_1^x, 
g_2^y, \pairing(g_1,g_2)^{z}\right)
\end{array}$$ are computationally 
indistinguishable for random chosen $g_1, g_2,x,y,z,\alpha$.
\end{theorem}

\noindent
{\bf Proof.} We use a random reduction. For a contradiction, 
assume that there is a polynomial time probabilistic 
algorithm $\D$ that distinguishes $\X$ and $\Y$ with a 
non-negligible probability $\delta_k$. We construct a polynomial 
time probabilistic algorithm $\A$ that distinguishes 
$(\R, \pairing(g,g)^{t})$ and $(\R, \pairing(g,g)^{uvw})$ with 
$\delta_k$, where $\R=(g,g^u,g^v,g^w)$ and $u,v,w,t$ are uniformly 
at random in $Z_q$.
Let the input of $\A$ be $(\R, \pairing(g,g)^{\tilde{t}})$, 
where $\tilde{t}$ is either ${uvw}$ or uniformly at random 
in $Z_q$. We construct $\A$ as follows.
$\A$ chooses random $c_1,c_2,c_3,c_4,c_5\in Z_q$ and sets
$g_1=g^{c_1}$, $g_2=g^{c_2}$, 
$g_1^\alpha=g^{uc_1c_3}$, $g_2^\alpha=g^{uc_2c_3}$, 
$g_1^x=g^{vc_1c_4}$, $g_2^y=g^{wc_2c_5}$, 
and 
$\pairing(g_1, g_2)^{\tilde{z}}=
\pairing(g, g)^{\tilde{t}c_1c_2 c_3c_4c_5}$.
Let $\A\left(\R, \pairing(g,g)^{\tilde{t}}\right)=
\D\left(g_1, g_2,g_1^\alpha, g_2^\alpha,g_1^x, 
g_2^y, \pairing(g_1,g_2)^{\tilde{z}}\right).$
Note that if $\tilde{t}=uvw$, then $c_1,c_2,\alpha,x,y$ are uniform
in $Z_q$ (and independent of each other and of $u,v,w$) 
and $xy\alpha=\tilde{z}$. Otherwise, $c_1,c_2,\alpha,x,y$ are uniform
in $Z_q$ and independent of each other and of $u,v,w$. Therefore,
by the definitions, 
$$\begin{array}{ll}
&\Pr\left[\A\left(\R, \pairing(g,g)^{uvw}\right)=1\right]
=\Pr\left[\D(\X)=1\right]\\
\mbox{and}\quad\quad&\Pr\left[\A\left(\R,\pairing(g,g)^{t}\right)=1\right]
=\Pr\left[\D(\Y)=1\right]
\end{array}$$
Thus $\A$ distinguishes $\langle g, g^u, g^v, g^w, \pairing(g,g)^{t}\rangle$
and $\langle g, g^u, g^v, g^w, \pairing(g,g)^{uvw}\rangle$ with 
$\delta_k$. This is a contradiction.
\hfill$\Box$

\vskip 3pt
Though Theorem \ref{pfsthm} shows that the protocol IDAK 
achieves weak perfect forward secrecy even if both participating
parties' long term private keys were corrupted, IDAK does not have perfect
forward secrecy when the master secret $\alpha$ were leaked. 
The perfect forward secrecy against the corruption of $\alpha$ 
could be achieved by requiring Bob (the responder in the IDAK protocol)
to send $g_{\ID_A}^y$ in addition to the value 
$R_B=g_{\ID_B}^y$
and by requiring both parties to compute the shared secret
as $H(g_{\ID_A}^{xy}||sk_{AB})$ where $sk_{AB}$ is the shared secret
established by the IDAK protocol.

\section{Key compromise impersonation (KCI) resilience}
\label{kcridak}
In this section, we informally show that the protocol IDAK has the 
key compromise impersonation resilience property.
That is, if Alice loses her private key $d_A=g_{ID_A}^\alpha$,
then the adversary still could not impersonate Bob to 
Alice.  For a formaly proof of KCI, we still need to consider the
information obtained by the adversary by {\bf Reveal}, {\bf Extract}, 
{\bf Send}, {\bf Corrupt} queries in other sessions. This
will be done in the final version of this paper.

In order to show KCI for IDAK,  it is (informally) sufficient to show that
the two distributions $\left(\R, \pairing\left(g_{\ID_A}^x\cdot 
g_{\ID_A}^{\pi(g_{\ID_A}^x, R_B)}, R_B\cdot g_{\ID_B}^{\pi(R_B,g_{\ID_A}^x)}
\right)^\alpha\right)$ and 
$\left(\R, \pairing(g_{\ID_A},g_{\ID_B})^{z}\right)$ 
are computationally indistinguishable for 
$\R=(g_{\ID_A}^\alpha,g_{\ID_A}^x,R_B)$,
where $g_{\ID_A}, g_{\ID_B},x,z,\alpha$ are chosen uniform at random,
and $R_B$ is chosen according to some probabilistic polynomial time
distribution. Since the value 
$\pairing\left(g_{\ID_A}^{\pi(g_{\ID_A}^x, R_B)}, 
R_B\cdot g_{\ID_B}^{\pi(R_B,g_{\ID_A}^x)}
\right)^\alpha$ is known, 
it is sufficient to prove the following theorem.

\begin{theorem}
\label{kcrthm}
Let $\group=\{G_\n\}$ be a bilinear group family,
$G_\n=\langle G, G_1, \pairing\rangle$. 
Assume that DBDH-Assumption holds for $\group$.
Then the two distributions
$$\begin{array}{ll}
&\X=\left(g_1, g_2,g_1^\alpha, g_1^x,
R_B, \pairing\left(g_1^x, R_B\cdot g_2^{\pi(R_B,g_1^x)}
\right)^\alpha\right)\\
\mbox{and}\quad\quad &
\Y=\left(g_1, g_2,g_1^\alpha,g_1^x, 
R_B, \pairing(g_1,g_2)^{z}\right)
\end{array}$$ are computationally 
indistinguishable for random chosen $g_1, g_2,x,z,\alpha$,
where $R_B$ is chosen according to some probabilistic polynomial time
distribution.
\end{theorem}

\noindent
{\bf Proof.} Since $g_1^x$ is chosen uniform at random,
and $\pi$ is a random oracle, we may assume that 
$R_B\cdot g_2^{\pi(R_B,g_1^x)}$ is uniformly distributed over
$G$ when $R_B$ is chosen according to any probabilistic polynomial time
distribution. Thus the proof is similar to the proof of 
Theorem \ref{pfsthm} and the details are omitted.
The theorem could also be proved using the Splitting lemma 
\cite{fork} which was used to prove the fork lemma.
Briefly, the Splitting lemma translates the fact that when a subset
$A$ is ``large'' in a product space $X\times Y$, it has many large sections.
Using the Splitting lemma, one can show that if $\D$ can distinguish
$\X$ and $\Y$, then by replaying $\D$ with different random oracle
$\pi$, one can get sufficient many tuples $(g_1,g_2,g_1^\alpha,g_1^x,R_B,
\pi_1,\pi_2)$
such that (1) $\pi_1(R_B,g_1^x)\not \pi_2(R_B,g_1^x)$;  
(2) $\D$ distinguishes $\X_1$ and $\Y$ (respectively $\X_2$ and $\Y$) 
when $z$ is uniformly chosen but other values takes the values from 
the above tuple with $\pi_1$ (respectively $\pi_2$). 
Since  $\pairing\left(g_1^x, R_B\cdot g_2^{\pi_1(R_B,g_1^x)}
\right)^\alpha/\pairing\left(g_1^x, R_B\cdot g_2^{\pi_2(R_B,g_1^x)}
\right)^\alpha=\pairing\left(g_1, g_2
\right)^{x\alpha ({\pi_1(R_B,g_1^x)}-{\pi_2(R_B,g_1^x)})}$. Thus,
for the above tuple, we can distinguish 
$\pairing\left(g_1, g_2\right)^{x\alpha }$ from 
$\pairing\left(g, g\right)^{z}$ for random chosen $z$.
This is a contradiction with the DBDH-Assumption.
\hfill$\Box$


\section{Appendix}
\subsection{Proof of Theorem \ref{perfectcbdh}}

The fact that the CBDH-Assumption implies the perfect-CBDH-Assumption
is trivial. The converse is proved by the self-random-reduction technique
(see \cite{bmicali,naor}). Let $\oracle$ be a CBDH oracle. 
That is, there exists a ${c}>0$ such that (\ref{cbdhe}) holds with 
$\C$ replaced with $\oracle$. We construct a perfect-CBDH algorithm
$\C$ which makes use of the oracle $\oracle$. Given
$g, g^x,g^y,g^z\in G$, algorithm $\C$ must compute $\pairing(g,g)^{xyz}$ 
with overwhelming probability. Consider the following algorithm: 
select $a,b,c\in_R Z_q$ (unless stated explicitly, 
we use $x\in_R X$ to denote that $x$ is randomly chosen 
from $X$ in the remainder of this paper) and output
$$I_{x,y,z,a,b,c}=\oracle(g,g^{x+a},g^{y+b}, g^{z+c})
\cdot \pairing(g,g)^{-(abz+abc+ayz+ayc+xbz+xbc+xyc)}.$$
One can easily verify that if $\oracle(\n,g,g^{x+a},g^{y+b}, g^{z+c})=
\pairing(g,g)^{(x+a)(y+b)(z+c)}$, then $I_{x,y,z,a,b,c}=
\pairing(g,g)^{xyz}$.
Consequently, standard amplification techniques can be used to construct 
the algorithm $\C$. The details are omitted.

\subsection{Proof of Lemma \ref{secondlemma}}
For a contradiction, assume that there is a 
probabilistic polynomial-time algorithm $\D$ that distinguishes the
two distributions $\X_2$ and $\Y_2$ with non-negligible probability
$\delta_k$. In the following we construct a probabilistic 
polynomial-time algorithm $\D^\prime$ to distinguish the
two distributions $\X_1$ and $\Y_1$.
$\D^\prime$ is defined by letting
$\D^\prime\left(\R,X, Y\right)=\D\left(\R,
X\cdot Y\right)$ for all $\R$, and $X,Y\in G_1$. 
By this definition, we have 
$\pr\left[\D^\prime_r(\X_1)=1|\R,r\right] 
=\pr\left[\D_r(\X_2)=1|\R,r\right]$, for 
any fixed internal coin tosses $r$ of $\D$ and $\D^\prime$.

Let $D_{\R,r}^{\D} = 
\left\{X: \D_r\left(\R, X\right)=1\right\} \mbox{ and }
D_{\R,r}^{\D^\prime} = 
\left\{(X,Y): \D^\prime_r\left(\R,
X, Y\right)=1\right\}$.
By definition of $\D^\prime$, we have 
$D_{\R,r}^{\D^\prime}=\{(X,Y): X\cdot Y\in D_{\R,r}^{\D}\}.$
It follows that 
$|D_{\R,r}^{\D^\prime}|=q|D_{\R,r}^{\D}|$ and 
$\pr\left[\D^\prime_r(\Y_1)=1|\R,r\right] 
={|D_{\R,r}^{\D^\prime}|}/{q^2}
={|D_{\R,r}^{\D}|}/{q} 
=\pr\left[\D_r(\Y_2)=1|\R,r\right]$.
Thus we have
\begingroup
\def\arraystretch{1.5}  
$$\begin{array}{ll}
&\left|\pr\left[\D^\prime\left(\X_1\right)=1\right]
-\pr\left[\D^\prime(\Y_1)=1\right]\right|\\
=&\left|\sum_{\R,r}\pr[\R,r]\cdot\left(
\pr\left[\D^\prime_r(\X_1)=1|\R,r\right]-
\pr\left[\D^\prime_r(\Y_1)=1|\R,r\right]\right)\right|\\
=&\left|\sum_{\R,r}\pr[\R,r]\cdot\left(
\pr\left[\D_r(\X_2)=1|\R,r\right]-
\pr\left[\D_r(\Y_2)=1|\R,r\right]\right)\right|\\
=&\left|\pr\left[\D(\X_2)=1\right]-
\pr\left[\D(\Y_2)=1\right]\right|\\
>&\delta_k.
\end{array}$$
\endgroup
Hence, $\D^\prime$ distinguishes the distributions
$\X_1$ and $\Y_1$ with non-negligible probability $\delta_k$. 
This contradicts the assumption of the Lemma.

\subsection{Proof of Lemma \ref{feedbackBDH}}
The Lemma could be proved using complicated version of 
the Splitting lemma 
by Pointcheval-Stern \cite{fork} (see the proof of Theorem \ref{kcridak}). 
In the following, we use the random reduction to prove the lemma.

1. For a contradiction, assume that there is a polynomial 
time probabilistic algorithm $\D$ 
that distinguishes $\X$ and $\Y$. We construct a polynomial 
time probabilistic algorithm $\A$ that distinguishes 
$\langle g,g^u,g^v,g^w$, 
$\pairing(g,g)^{a}\rangle$ and $\langle g,g^u,g^v,g^w, 
\pairing(g,g)^{uvw}\rangle$ with 
$\delta_k$, where $u,v,w,a$ are uniformly at random in $Z_q$.

Let the input of $\A$ be $\langle g,g^u,g^v,g^w, 
\pairing(g,g)^{\tilde{a}}\rangle$, where $\tilde{a}$ is either 
${uvw}$ or uniformly at random in $Z_q$. $\A$ chooses uniformly at random 
$c_1,c_2,c_3,x,x_0\in Z_q$, sets
$g^\alpha=g^{c_1u+c_2}$, $g^\beta=g^{v+c_3}$, $g^\gamma=g^{w+c_4}$, 
chooses uniformly at random $r\in Z_q$ or lets $g^r=g^\beta$,
chooses $g^{\gamma y_0},g_\A\in G$
within polynomial time according to any distribution given the view 
$({g^x,g^r,g^\alpha,g^\beta, g^\gamma,g^{\beta x_0}})$ 
(the distributions for $g_\A\in G$ and 
$g^{\gamma y_0}$ could be different).
Since $g^x$ and $g^{\beta x_0}$ are uniformly chosen from $G$,
we may assume that the values of $\pi(g^x,g_\A)$ and
$\pi(g^{\gamma y_0},g^{\beta x_0})$ are unknown yet.
Without loss of generality, we may assume that $x+\beta \pi(g^x,g_\A)$
and $y_0+\pi(g^{\gamma y_0},g^{\beta x_0})$
take values $c_5$ and $c_6$ respectively,
where $c_5$ and $c_6$ are uniformly chosen from $Z_q$. In a summary,
the value of $\R$ could be computed from $g^u,g^v,g^w,c_1,c_2,c_3,c_4,c_5$
efficiently. $\A$ then sets 
$$\pairing(g,g)^{\tilde{t}}=\pairing(g,g)^{c_1\tilde{a}+c_4(c_1u+c_2)(v+c_3)
+w(c_1uc_3+c_1v+c_2c_3)}.$$
$\A$ can compute
$\pairing(g,g)^{(x_0+\pi(g^{\beta x_0},g^{\gamma y_0}))
(y_0+\pi(g^{\gamma y_0},g^{\beta x_0}))\tilde{t}}$
using the values of $\pairing(g,g)^{\tilde{t}}$, $x_0$, 
$\pi(g^{\beta x_0}, g^{\gamma y_0})$, $c_6$.
Let $\A\left(g,g^u,g^v,g^w,\pairing(g,g)^{\tilde{a}}\right)=
\D(\tilde{\X})$, where $\tilde{\X}$ is obtained from
$\Y$ by replacing $t$ with $\tilde{t}$ and taking the remaining 
values as defined above.

Note that if $\tilde{a}=uvw$, then $\tilde{t}=\alpha\beta\gamma$,
and $\tilde{\X}$ is distributed according to the distribution $\X$. 
That is, $\alpha, \beta, \gamma, x, x_0$ are uniform
in $Z_q$ and independent of each other and of $(u,v,w)$,
($r$, $g_\A$, $g^{\gamma y_0}$) is chosen according to the 
specified distributions without violating DBDH-Assumption. 
Otherwise, $\tilde{\X}$ is distributed 
according to the distribution $\X$, and $\tilde{t}$ is uniform 
in $Z_q$ and independent of $\alpha, \beta, \gamma, x, x_0,
r,u,v,w, g_\A,g^{\gamma y_0}$. Therefore, by definitions, 
$$\begin{array}{ll}
&\Pr\left[\A\left(g,g^u,g^v,g^w,\pairing(g,g)^{uvw}\right)=1\right]
=\Pr\left[\D(\X)=1\right]\\
\mbox{and}\quad\quad&
\Pr\left[\A\left(g,g^u,g^v,g^w,\pairing(g,g)^{a}\right)=1\right]
=\Pr\left[\D(\Y)=1\right]
\end{array}$$
Thus $\A$ distinguishes 
$\langle g,g^u,g^v,g^w, \pairing(g,g)^{a}\rangle$ and $\langle g,g^u,g^v,g^w, 
\pairing(g,g)^{uvw}\rangle$ with 
$\delta_k$, where $a$ is uniform at random in $Z_q$. This is a contradiction.

2. This part of the Lemma could be proved in the same way. The details
are omitted.

3.  Since ``$g_{\A, i}\not=g^{\beta x_0}$ or $g_{\A,l}\not=g^{\gamma y_0}$'',
we may assume that the values of $\pi(g_{\A, i},g_{\A,l})$ and
$\pi(g_{\A, l},g_{\A,i})$ are unknown yet. By the random oracle property of
$\pi$, this part of the Lemma could be proved in the same way as in item 1. 
The details are omitted.

\section{Proof of Theorem \ref{securityprooftheorem}}

\noindent
{\bf Proof.} By Theorem \ref{passiverandom},
the condition 1 in the Definition \ref{keysecuredef}
is satisfied for the IDAK key agreement protocol. In the following,
we show that the condition 2 is also satisfied.

For a contradiction, assume that the adversary $\A$ has non-negligible
advantage $\delta_k=\mathrm{Adv}^{\A}(k)$ in guessing the value of $b$ 
after the {\bf Test} query. We show how to construct a simulator 
$\Simu$ that uses $\A$ as an oracle to distinguish 
the distributions $\X$ and $\Y$ in the item 3 of Lemma 
\ref{feedbackBDH} with non-negligible advantage 
$2\delta_k(q_E-2)^2/q_E^4$, where $q_E$ denotes
the number of distinct $H$-{\bf queries} that the algorithm $\A$ has made.
The game between the challenger and the simulator $\Simu$ starts with
the challenger first generating bilinear groups 
$G_\n=\langle G, G_1, \pairing\rangle$ by running the algorithm
{\bf Instance Generator}. The challenger then chooses
$\alpha,\beta,\gamma,t\in_R Z_q$ and $b\in_R\{0,1\}$. The challenger
gives the tuple $\langle \n, g, g^\alpha,g^\beta,g^\gamma, 
\pairing(g,g)^{\tilde{t}}\rangle$
to the algorithm $\Simu$ where $\tilde{t}={\alpha\beta\gamma}$ if $b=1$ and 
$\tilde{t}=t$ otherwise. During the simulation, the algorithm 
$\Simu$ can ask the challenger to provide randomly chosen 
$g^{x_i}$. $\Simu$ may then choose (with the help
of $\A$ perhaps) $g_{\A,l}$ 
within polynomial time according to any distribution given the view 
$({g^{x_i},g^{r_j},g^\alpha, g^\beta,g^\gamma, g^{\alpha x_0}: 
i,j,l\le m})$ and sends  $g_{\A,l}$ to the challenger. The 
challenger responds with 
$\pairing(g^{x_i+\beta\pi(g^{x_i},g_{\A,l})},
g_{\A,l}\cdot g^{r_j\pi(g_{\A,l},g^{x_i})})^\alpha$.
At the end of the simulation, the algorithm $\Simu$ is supposed to 
output its guess $b'\in\{0,1\}$ for $b$.
It should be noted that if $b=1$, then the output of the challenger 
together with the values $g_{\A,l}$ selected by the simulator
$\Simu$ is the tuple $\X$ of Lemma \ref{feedbackBDH}, 
and is the tuple $\Y$ of Lemma \ref{feedbackBDH} if $b=0$.
Thus the simulator $\Simu$ could be used to distinguish 
$\X$ and $\Y$ of Lemma \ref{feedbackBDH}.

The algorithm $\Simu$ selects two integers $I,J\le q_E$ randomly 
and works by interacting with $\A$ as follows:

\vskip 3pt
\noindent
{\bf Setup:} Algorithm $\Simu$ gives $\A$ the IDAK system parameters
$\langle q, G, G_1, \pairing, H,\pi\rangle$ where $q, G, G_1,\pairing$
are parameters from the challenger, $H$ and $\pi$ are random oracles 
controlled by $\Simu$ as follows.

\noindent
$H$-{\bf queries}: At any time algorithm $\A$ can query the random oracle
$H$ using the queries ${\bf Extract}(\ID_i)$ or 
${\bf GetID}(\ID_i)=H(\ID_i)$. 
To respond to these queries algorithm $\Simu$ maintains an $H^{list}$
that contains a list of tuples 
$\langle \ID_i, g_{\ID_i}\rangle$. The list is initially
empty. When $\A$ queries the oracle $H$ at a point $\ID_i$, $\Simu$
responds as follows:
\begin{enumerate}
\item If the query $\ID_i$ appears on the  $H^{list}$
in a tuple $\langle \ID_i, g_{\ID_i}\rangle$, then 
$\Simu$ responds with $H(\ID_i)=g_{\ID_i}$.
\item Otherwise, if this is the $I$-th new query of the random oracle
$H$, $\Simu$ responds with $g_{\ID_i}=H(\ID_i)=g^{\beta}$, 
and adds the tuple $\langle \ID_i, g^\beta\rangle$ to the  $H^{list}$.
If this is the $J$-th new query of the random oracle,
$\Simu$ responds with $g_{\ID_i}=H(\ID_i)=g^{\gamma}$, 
and adds the tuple $\langle \ID_i, g^\gamma\rangle$ to the  $H^{list}$.
\item In the remaining case, $\Simu$ selects a random $r_i\in Z_q$, 
responds with $g_{\ID_i}=H(\ID_i)=g^{r_i}$, 
and adds the tuple $\langle \ID_i, g^{r_i}\rangle$ to the 
$H^{list}$.
\end{enumerate}

\noindent
$\pi$-{\bf queries}: At any time the challenger, the algorithm $\A$, and
the algorithm $\Simu$ can query the random oracle
$\pi$. To respond to these queries algorithm $\Simu$ maintains a $\pi^{list}$
that contains a list of tuples 
$\langle g_1,g_2, \pi(g_1,g_2)\rangle$. The list is initially
empty. When $\A$ queries the oracle $\pi$ at a point $(g_1,g_2)$, $\Simu$
responds as follows:
If the query $(g_1,g_2)$ appears on the  $\pi^{list}$
in a tuple $\langle (g_1,g_2), \pi(g_1,g_2)\rangle$, then 
$\Simu$ responds with $\pi({g_1,g_2})$.
Otherwise, $\Simu$ selects a random $v_i\in Z_q$, 
responds with $\pi(g_1,g_2)=v_i$, and adds the tuple 
$\langle (g_1,g_2), {v_i}\rangle$ to the $\pi^{list}$.
Technically, the random oracle $\pi$ could be held by an independent
third party to avoid the confusion that the challenger also needs to 
access this random oracle also.

\vskip 3pt
\noindent
{\bf Query phase:} $\Simu$ responds to 
$\A$'s queries as follows. 

For a ${\bf GetID}(\ID_i)$ query, $\Simu$ runs the $H$-{\bf queries} 
to obtain a $g_{\ID_i}$ such that $H(\ID_i)=g_{\ID_i}$, and responds
with $g_{\ID_i}$.

For an ${\bf Extract}(\ID_i)$ query for the long term private key, 
if $i=I$ or $i=J$, then $\Simu$ reports failure and terminates. Otherwise,
$\Simu$ runs the $H$-{\bf queries} to obtain $g_{\ID_i}=H(\ID_i)=g^{r_i}$, 
and responds $d_{\ID_i}=\left(g^\alpha\right)^{r_i}=g_{\ID_i}^\alpha$.

For a ${\bf Send}(\Pi_{i,j}^s, X)$ query, we distinguish the following
three cases:
\begin{enumerate}
\item $X=\lambda$. If $i=I$ or $J$, $\Simu$ asks the challenger 
for a random $R_i\in G$ (note that $\Simu$ does not know 
the discrete logarithm of $R_i$ with base $g_{\ID_i}$), 
otherwise $\Simu$ chooses a random $u_i\in Z_q^*$ and 
sets $R_i=g_{\ID_i}^{u_i}$. $\Simu$ lets $\Pi^s_{i,j}$ reply
with $R_i$. That is, we assume that $\ID_i$ is carrying out an
IDAK key agreement protocol with $\ID_j$ and $\ID_i$ sends
the first message $R_i$ to $\ID_j$.
\item $X\not=\lambda$ and the transcript of the oracle $\Pi_{i,j}^s$ 
is empty. In this case, $\Pi_{i,j}^s$ is the responder to the protocol 
and has not sent out any message yet. 
If $i=I$ or $J$, $\Simu$ asks the challenger for a random $R_i\in G$, 
otherwise 
$\Simu$ chooses a random $u_i\in Z_q^*$ and sets $R_i=g_{\ID_i}^{u_i}$. 
$\Simu$ lets $\Pi^s_{i,j}$ reply with $R_i$ and marks 
the oracle $\Pi_{i,j}^s$ as completed.
\item  $X\not=\lambda$ and the transcript of the oracle $\Pi_{i,j}^s$ 
is not empty. In this case, $\Pi_{i,j}^s$ is the protocol initiator
and should have sent out the first message already. 
Thus $\Pi_{i,j}^s$ does not need to respond anything.
After processing the query ${\bf Send}(\Pi_{i,j}^s, X)$, $\Simu$ marks 
the oracle $\Pi_{i,j}^s$ as completed.
\end{enumerate}

For a ${\bf Reveal}(\Pi_{i,j}^s)$ query, if $i\not= I$ and $i\not= J$,
$\Simu$ computes the session key 
$sk_{ij}=\pairing(g_{\ID_j}^{\pi(R_j,R_i)}\cdot 
R_j$, $d_{\ID_i}^{(u_i+\pi(R_i,R_j))})$ and responds with
$sk_{ij}$, here $R_j$ is the 
message received by $\Pi_{i,j}^s$. Note that the message $R_j$ 
may not necessarily be sent by the oracle  $\Pi_{j,i}^{s'}$ for some
$s'$ since it could have been a bogus message from $\A$.
Otherwise, $i=I$ or $i=J$. Without loss of generality, we assume that 
$i=I$. In this case, the oracle  $\Pi_{I,j}^s$ dose not know its
private key $g^{\beta\alpha}$. Thus it needs help from the challenger
to compute the shared session key. Let $R_I$ and $R_j$ be the messages
that $\Pi_{I,j}^s$  has sent out and received respectively.
$\Pi_{I,j}^s$ gives these two values to the challenger and the challenger
computes the shared session key 
$sk_{Ij}=\pairing\left(g_{\ID_j}^{\pi(R_j,R_i)}\cdot 
R_j, R_I^{\alpha h} g^{\pi(R_I,R_j)\alpha\beta }\right)$.
$\Pi_{I,j}^s$ then responds with $k_{Ij}$. 

For a ${\bf Corrupt}(i)$ query, if $i=I$ or $i=J$, then $\Simu$ 
reports failure and terminates. Otherwise, $\Simu$ responds with 
$d_{{\ID}_i}=\left(g^\alpha\right)^{r_i}=g_{\ID_i}^\alpha$.

For the {\bf Test}$(\Pi_{i,j}^s)$ query, if $i\not=I$ or $j\not=J$,
then $\Simu$ reports failure and terminates. Otherwise, assume that
$i=I$ and $j=J$. Let $R_I=g_{\ID_I}^{u_I}$ be the message 
that $\Pi_{i,j}^s$ sends out (note that the challenger generated this
message) and $R_J=g_{\ID_J}^{u_J}$ be the 
message that $\Pi_{i,j}^s$ receives (note that $R_J$ could be the message
that the challenger generated or could be generated by the algorithm $\A$).
$\Simu$ gives the messages $R_I$ and $R_J$ to the challenger. The challenger 
computes $X=\pairing(g,g)^{(u_I+\pi(R_I,R_J))(u_J+\pi(R_J,R_I))\tilde{t}}$ 
and gives $X$ to $\Simu$. $\Simu$ responds with $X$. 
Note that if $\tilde{t}=\alpha\beta\gamma$, then $X$ is the 
session key. Otherwise, $X$ is a uniformly distributed group element.

\vskip 3pt
\noindent
{\bf Guess:} After the {\bf Test}$(\Pi_{i,j}^s)$ query, 
the algorithm $\A$ may issue other queries before finally
outputs its guess $b'\in\{0,1\}$. 
Algorithm $\Simu$ outputs $b'$ as its guess to the challenger.

\vskip 3pt
\noindent
{\bf Claim:} If $\Simu$ does not abort during the simulation then
$\A$'s view is identical to its view in the real attack. Furthermore,
if $\Simu$ does not abort, then $\left|\Pr[b=b']-\frac{1}{2}\right|
>\delta_k$, where the probability is over all random coins used
by $\Simu$ and $\A$.

\vskip 3pt
\noindent
{\em Proof of Claim:} The responses to $H$-{\bf queries} 
and $\pi$-{\bf queries} are the same as in the real attack since 
the response is uniformly distributed. All responses to the 
getID queries, private key extract queries,
message delivery queries, reveal queries, and corrupt queries
are valid. It remains to show that the response to the test query is
valid also. When $\tilde{t}$ is uniformly distributed over $Z_q$, 
then Theorem \ref{passiverandom} shows
that $X=\pairing(g,g)^{(u_I+\pi(R_I,R_J))(u_J+\pi(R_J,R_I))\tilde{t}}$ 
is uniformly distributed 
over $G$ and is computationally indistinguishable from a random 
value before $\A$'s view. Therefore, by definition of the algorithm 
$\A$, we have $\left|\Pr[b=b']-\frac{1}{2}\right|>\delta_k$.
\hfil$\Box$

\vskip 3pt
Suppose $\A$ makes a total of $q_E$ $H$-queries.
We next calculate the probability that $\Simu$ does not abort 
during the simulation. The probability that 
$\Simu$ does not abort for {\bf Extract} queries is 
$(q_E-2)/q_E$. The probability that $\Simu$ does not abort 
for {\bf Corrupt} queries is $(q_E-2)/q_E$. The probability that 
$\Simu$ does not abort for {\bf Test} queries is 
$2/q_E^2$. Therefore, the probability that 
$\Simu$ does not abort during the simulation is $2(q_E-2)^2/q_E^4$.
This shows that $\Simu$'s advantage in distinguishing 
the distributions $\X$ and $\Y$ in Lemma \ref{feedbackBDH}
is at least $2\delta_k(q_E-2)^2/q_E^4$ which is non-negligible.

To complete the proof of Theorem \ref{securityprooftheorem}, it remains
to show that the communications between $\Simu$ and the challenger
are carried out according to the distributions $\X$ and $\Y$ of 
Lemma \ref{feedbackBDH}. For a ${\bf Reveal}(\Pi_{I,j}^s)$  query,
the challenger outputs $\pairing\left(g_{\ID_j}^{\pi(R_j,R_I)}\cdot 
R_j, R_I^{\alpha h} g^{\pi(R_I,R_j)\alpha\beta }\right)$ to the 
algorithm $\Simu$. Let $R_I=g^{x}$, $R_j=g_\A$, and 
$g_{\ID_j}=g^r$. Then $x$ is chosen uniform at random from $Z_q$, 
$r$ is chosen uniform at random from $Z_q^*$ when $j\not= J$ 
or $r=\gamma$ when $j=J$, and the value of $g_\A$ is chosen by the algorithm
$\A$ or by the algorithm $\Simu$ or by the challenger in probabilistic
polynomial time according to the current views. 
For example, if $g_\A$ is chosen by the algorithm $\A$, then $\A$ may
generate $g_\A$ as the combination (e.g., multiplication) 
of some previously observed messages/values or generate it randomly. 
Thus the communication between the challenger and the algorithm $\Simu$ 
during ${\bf Reveal}(\Pi_{I,j}^s)$ queries is carried out according
to the distributions $\X$ and $\Y$ of Lemma \ref{feedbackBDH}.
The case for ${\bf Reveal}(\Pi_{J,j}^s)$ queries is the same.

For the {\bf Test}$(\Pi_{I,J}^s)$ query, the challenger outputs 
$X=\pairing(g,g)^{(u_I+\pi(R_I,R_J))(u_J+\pi(R_J,R_I))\tilde{t}}$ to 
the algorithm $\Simu$, where $R_I=g^{\beta u_I}$ and $R_J=g^{\gamma u_J}$. 
Let $x_0=u_I$ and $y_0=u_J$. Then $x_0$ is chosen uniform at random from $Z_q$
and the value of $g^{\gamma y_0}$ is chosen by the algorithm
$\A$ or by the challenger in probabilistic polynomial time 
according to the current views. Similarly, $\A$ may choose 
$g^{\gamma y_0}$ as the combination (e.g., multiplication) 
of some previously observed messages/values. 
The communication between the challenger and the 
algorithm $\Simu$ during the ${\bf Test}(\Pi_{I,J}^s)$ query is 
carried out according to the distributions $\X$ and $\Y$ 
of Lemma \ref{feedbackBDH}. 

It should be noted that after the {\bf Test}$(\Pi_{I,J}^s)$ query,
the adversary may create bogus oracles for the participants
$\ID_I$ and $\ID_J$ and send bogus messages that may depend
on all existing communicated messages (including messages 
held by the oracle $\Pi_{I,J}^s$) and then reveal session keys
from these oracles. 
In particular, the adversary may
play a man in the middle attack by modifying the messages
sent from $\Pi_{I,J}^s$ to $\Pi_{J,I}^{s'}$ and modifying 
the messages sent from $\Pi_{J,I}^{s'}$ to $\Pi_{I,J}^s$.
Then the oracles $\Pi_{J,I}^{s'}$ and $\Pi_{I,J}^s$ are not
matching oracles. Thus $\A$ can reveal the session key held by 
the oracle $\Pi_{J,I}^{s'}$
before the guess. In the $\R_2$ part in the distributions $\X$ and $\Y$ 
of Lemma \ref{feedbackBDH}, we have the condition 
``$g_{\A,i}\not=g^{\beta x_0}$ or $g_{\A,l}\not=g^{\gamma y_0}$''
(this condition holds since the algorithm $\A$ has not revealed
the matching oracles for $\Pi_{I,J}^s$).
If both $g_{\A,i}\not=g^{\beta x_0}$ and $g_{\A,l}\not=g^{\gamma y_0}$,
then the oracle $\Pi_{J,I}^{s'}$ is a matching oracle for $\Pi_{I,J}^s$
and $\A$ is not allowed to reveal the session key held by 
the oracle $\Pi_{J,I}^{s'}$.
Thus the communication between 
the challenger and the algorithm $\Simu$ during these 
${\bf Test}(\Pi_{I,J}^s)$ query is carried out according to 
the distributions $\X$ and $\Y$ of Lemma \ref{feedbackBDH}. 

In the summary, all communications between the challenger and $\Simu$
are carried out according to the distributions $\X$ and $\Y$ 
of Lemma \ref{feedbackBDH}. This completes the proof of the Theorem.
\hfill$\Box$

\section{Practical considerations and applications}
\label{idakpifunc}
\subsection{The function $\pi$}
Though in the security proof of IDAK key agreement protocol,
$\pi$ is considered as a random oracle. In practice, we can
use following simplified $\pi$ functions. 
\begin{itemize}
\item $\pi$ is a random oracle (secure hash function) from 
$G\times G$ to $Z_{2^{\lceil\log q\rceil/c}}^*$ (e.g., $c=2$).
\item If $g_1=(x_{g_1},y_{g_1}), g_2=(x_{g_2},y_{g_2})\in G$ are
points on an elliptic curve, then let
$\pi(g_1,g_2)=\bar{x}_g \mbox{ mod } 2^{|x_g|/2}$ where 
$\bar{x}_g=x_{g_1}\oplus x_{g_2}$. That is,
$\pi(g_1,g_2)$ is the exclusive-or of the second half parts of 
the first coordinates of the elliptic curve points $g_1$ and $g_2$. 
\item $\pi$ is a random oracle that the output only depends 
on the the first input variable or any of the above function
restricted in such a way that the output only depends 
on the the first input variable. In another word, 
$\pi:G\rightarrow Z_q^*$.
\end{itemize}
It should be noted any $\pi$ function, for which Lemma \ref{feedbackBDH} 
holds, can be used in the IDAK protocol. Though we do not know whether 
Lemma \ref{feedbackBDH} holds for $\pi$ functions that we have listed
above, we have strong evidence that this is true.
First, if we assume that the group $G_2$ is a generic group
in the sense of Nechaev \cite{nechaev} and Shoup \cite{shoupgm}. 
Then we can prove that Lemma \ref{feedbackBDH} holds for the above
$\pi$ functions. Secondly, if the distribution 
$\G({g^x,g^r,g^\alpha,g^\beta,g^\gamma,g^{\beta x_0}})$ 
in Lemma \ref{feedbackBDH} is restricted to the distribution: 
$$\{g^{f(x,r,\alpha, \beta, \gamma,\beta x_0,\mathbf{y})}: 
f \mbox{ is a linear function, }
\mathbf{y}\mbox{ is a tuple of uniformly random values from }Z_q\}.$$
Then we can prove that Lemma \ref{feedbackBDH} holds for the above
$\pi$ functions. We may conjecture that the adversary algorithm $\A$ 
can only generate $g_\A$ and $g^{\gamma y_0}$ 
according to the above distribution unless CDH-Assumption fails for 
$G$. Thus, under this conjecture (without the condition that $G_2$ is
a generic group), the above list of $\pi$ functions
can be used in IDAK protocol securely.

\subsection{Performance}
Our analysis in this section will be based on the assumption that 
$\pi$ is a random oracle (secure hash function) from 
$G\times G$ to $Z_{2^{\lceil\log q\rceil/2}}^*$.
Since the computational cost for Alice is the same as 
that for Bob. In the following, we will only analyze Alice's 
computation.

First, Alice needs to choose a random number $x$ and compute
$g_{\ID_A}^x$ in the group $G$.
In order for Alice to compute 
$sk=
\pairing\left(g_{\ID_B}^{s_B}\cdot 
R_B, g_{\ID_A}^{(x+s_A)\alpha}\right)$, she needs to do
$1.5$ exponentiation in $G$, one multiplication in $G$, and one pairing.
Thus in total, she needs to do $2.5$ exponentiation in $G$,
one multiplication in $G$, and one pairing.

Alternatively, Alice can compute the shared secret as
$sk=\pairing\left(g_{\ID_B}^{s_B}\cdot 
R_B, g_{\ID_A}^\alpha\right)^{(x+s_A)}$.
Thus for the entire IDAK protocol, Alice needs to do
$1.5$ exponentiation in $G$ (one for 
$g_{\ID_A}^x$ and $0.5$ for $g_{\ID_B}^{s_B}$), 
one multiplication in $G$, one pairing,
and one exponentiation in $G_1$.

The IDAK protocol could be sped up by letting each participant do some
pre-computation. For example, Alice can compute the values of 
$g_{\ID_A}^x$ and $g_{\ID_A}^{x\alpha}$ 
before the protocol session.
During the IDAK session, Alice can compute the shared secret as 
$sk=\pairing\left(g_{\ID_B}^{s_B}\cdot 
R_B, g_{\ID_A}^{x\alpha}\cdot  g_{\ID_A}^{\alpha s_A}\right)$
which needs $1$ exponentiation in $G$ ($0.5$ for
$g_{\ID_B}^{s_B}$ and $0.5$ for $ g_{\ID_A}^{\alpha s_A}$), 
$2$ multiplications in $G$, and one pairing.
Alternatively,  Alice can compute the shared secret as
$sk=\pairing\left(g_{\ID_B}^{s_B}\cdot 
R_B, g_{\ID_A}^{\alpha}\right)^{x+s_A}$
which needs $0.5$ exponentiation in $G$, 
one multiplication in $G$, one pairing,
and one exponentiation in $G_1$.
In a summary, Figure \ref{performancefigure} lists the computational
cost for Alice (an analysis of all other identity based key agreement
protocols shows IDAK is the most efficient one, details will be
given in the final version of this paper).
\begin{figure}[h]
\begin{center}
\begin{tabular}{|c|c|c|c|c|}\hline
 &\multicolumn{2}{|c|}{without pre-computation} &
              \multicolumn{2}{|c|}{with pre-computation} \\ \hline
& choice 1 & choice 2 & choice 1 & choice 2\\ \hline
pairing & 1 & 1 & 1 & 1\\ \hline
exponentiation in $G$ & 2.5 & 1.5 & 1 & 0.5 \\ \hline
multiplication in $G$ & 1 & 1 & 2 & 1\\ \hline
exponentiation in $G_1$& 0 & 1 & 0 & 1\\ \hline
\end{tabular}
\caption{IDAK Computational Cost for Alice}
\label{performancefigure}
\end{center}
\end{figure}


\begin{thebibliography}{99}
\bibitem{macbib}
M.~Bellare, R.~Canetti, and H.~Krawczyk.
Keying hash functions for message authentication. In:
{\em Advances in Cryptology, Crypto 96}, pages 1--15, 1996.

\bibitem{bck}
M.~Bellare, R.~Canetti, and H.~Krawczyk.
A modular approach to the design and analysis of authentication
and key exchange protocols. In:
{\em 30th Annual ACM Symposium on Theory of Computing}, 1998.

\bibitem{br}
M.~Bellare and P.~Rogaway.
Random oracles are practical: a paradigms for designing efficient
protocols.
In: {\em  Proc. 1st ACM Conference on Computer 
Communication Security}, pages 62--73, ACM Press, 1993.

\bibitem{br1}
M.~Bellare and P.~Rogaway.
Entity authentication and key distribution.
In: {\em Advances in Cryptology, Crypto 93},
LNCS 773 (1993), 232--249.


\bibitem{bmicali}
M.~Blum and S.Micali. How to generate cryptographically
strong sequence of pseudo-random bits. 
{\em SIAM J. Comput.} {\bf 13}:850--864, 1984.

\bibitem{boneh}
D.~Boneh.
The decision Diffie-Hellman problem.
In: {\em ANTS-III}, LNCS 1423 (1998), 48--63.

\bibitem{ibe}
D.~Boneh and M.~Franklin.
Identity-based encryption from the Weil pairing. 
{\em SIAM J. Computing} {\bf 32}(3):586--615, 2003.

\bibitem{canetti}
R.~Canetti. Universally composable security: a new paradigm for
cryptographic protocols. In: {\em Proc. 42nd FOCS}, 2001.

\bibitem{ck}
R.~Canetti and H.~Krawczyk.
Analysis of key-exchange protocols and their use for
building secure channels.  In:
{\em Advances in Cryptology, Eurocrypt 01},
LNCS 2045 (2001), 453--474. Full version available from
Cryptology ePrint Archive 2001-040 (\url{http://eprint.iacr.org/}).

\bibitem{ck02}
R.~Canetti and H.~Krawczyk.
Universally composable notions of key exchange and secure channels.
In: {\em Eurocrypt 02}.

\bibitem{chen}
L.~Chen and C.~Kudla.
Identity based authenticated key agreement protocols from pairing. 
In: {\em Proc. 16th IEEE Security Foundations Workshop}, pages 219--233.
IEEE Computer Society Press, 2003.

\bibitem{chengchen}
Z.~Cheng and L.~Chen.
On the security proof of McCullagh-Barreto's key
agreement protocol and its variants.
\url{http://eprint.iacr.org/2005/201.pdf}

\bibitem{cheng}
Z.~Cheng, M.~Nistazakis, R.~Comley, and L.~Vasiu. 
On indistinguishability-based security model of key agreement 
protocols-simple cases. In {\em Proc. of ACNS 04}, June 2004. 

\bibitem{kwang}
K.~Choo.
Revisit of McCullagh-Barreto two party ID-based authentication key
agreement protocols. 
\url{http://eprint.iacr.org/2004/343.pdf}

\bibitem{dh}
W.~Diffie and M.~Hellman.
New directions in cryptography.
{\em IEEE Transactions on Information Theory},
{\bf 6}(1976), 644--654.


\bibitem{fs}
A.~Fiat and A.~Shamir.
How to prove yourself: practical solutions of identification and signature
problems. In:
{\em Advances in Cryptology, Crypto 86},
LNCS 263 (1987), 186--194.

\bibitem{gp90}
M.~Girault and J.~Pailles.
An identity-based scheme providing zero-knowledge authentication and 
authenticated key exchange. In: {\em Proc. ESORICS 90}, pages 173--184.

\bibitem{joux}
A.~Joux. A one round protocol for tripartite Diffie-Hellman.
In: {\em Algorithmic number theory symposium, ANTS-IV}, LNCS 1838,
pages 385--394, 2000.


\bibitem{hmqv}
H.~Krawczyk. HMQV: a high-performance secure Diffie-Hellman
protocol. In: {\em Proc. Crypto 05}, Springer, 2005.

\bibitem{mqv}
L.~Law, A.~Menezes, M.~Qu, J.~Solinas, and S.~Vanstone.
An efficient protocol for authenticated key agreement.
{\em Designs, Codes and Cryptography}, {\bf 28}(2):119--134.

\bibitem{huawei}
S.~Li, Q.~Yuan, and J.~Li.
Towards security two-part authenticated key agreement protocols.
\url{http://eprint.iacr.org/2005/300.pdf}.


\bibitem{mccullagh}
P.~McCullagh and P.~Barreto.
A new two-party identity-based authenticated key 
agreement. {\em Proc. of CT-RSA 2005}, pages 262-274, LNCS 3376, Springer
Verlag, 2005.

\bibitem{mccullaghr}
P.~McCullagh and P.~Barreto.
A new two-party identity-based authenticated key 
agreement. \url{http://eprint.iacr.org/2004/122.pdf}

\bibitem{naor}
M.~Naor and O.~Reingold.
Number-theoretic constructions of efficient pseudo-random functions.
In: {\em 38th Annual Symposium on Foundations of Computer Science},
IEEE Press, 1998.

\bibitem{nechaev}
V.~Nechaev.
Complexity of a determinate algorithm for the discrete logarithm.
{\em Mathematical Notes}, {\bf 55}(1994), 165--172.

\bibitem{fips}
NIST Special Publication 800-56: Recommendation on key establishment
schemes, draft 2.0, 2003. 
\url{http://csrc.nist.gov/CryptoToolkit/kms/keyschemes-Jan03.pdf}.

\bibitem{ok86}
E.~Okamoto. Proposal for identity-based key distribution system.
{\em Electronics Letters} {\bf 22}:1283--1284, 1986.

\bibitem{fork}
D.~Pointcheval and J.~Stern.
Security arguments for digital signatures and blind signatures.
{\em J. Cryptology} {\bf 13}(3):361--396, 2000.

\bibitem{ryu}
E.~Ryu, E.~Yoon, and K.~Yoo. An efficient ID-based authenticated 
key agreement protocol from pairing. In:
{\em Networking 2004}, pages 1458--1463, LNCS 3042, Springer Verlag, 2004.

\bibitem{sok00}
R.~Sakai, K.~Ohgishi, and M.~Kasahara.
Cryptosystems based on pairing. In: {\em 2000 Symp. on Cryptography
and Information Security (SCIS 2000)}, Okinawa, Japan 2000.

\bibitem{scott}
M.~Scott. Authenticated ID-based key exchange and remote log-in
with insecure token and PIN number. 
\url{http://eprint.iacr.org/2002/164.pdf}


\bibitem{shamirid}
A.~Shamir. Identity-based cryptosystems and signature schemes.
In: {\em Advances in Cryptology, Crypto 84}, LNCS 196, pages 47--53,
Springer Verlag 1984.

\bibitem{shim}
K.~Shim. Efficient ID-based authenticated key agreement protocol
based on the Weil pairing.
{\em Electronics Letters} {\bf 39}(8):653--654, 2003.

\bibitem{shoupgm}
V.~Shoup. 
Lower bounds for discrete logarithms and related problems. In:
{\em Advances in Cryptology, Eurocrypt 97},
LNCS 1233 (1997), 256--266.

\bibitem{shoup}
V.~Shoup. On formal models for secure key exchange.
IBM Technical Report RZ 3120, 1999.

\bibitem{smart}
N.~P.~Smart. Identity-based authenticated key
agreement protocol based on Weil pairing. 
{\em Electronics Letters} {\bf 38}(13):630--632, 2002.

\bibitem{sun}
S.~Sun and B.~Hsieh.
Security analysis of Shim's authenticated key agreement protocols from
pairing.
\url{http://eprint.iacr.org/2003/113.pdf}

\bibitem{to91}
K.~Tanaka and E.~Okamoto.
Key distribution system for mail systems using ID-related information
directory. {\em Computers and Security} {\bf 10}:25--33, 1991.

\bibitem{xie1}
G.~Xie. Cryptanalysis of Noel McCullagh and Paulo S. L. M. Barreto's 
two-party identity-based key agreemenet. 
\url{http://eprint.iacr.org/2004/308.pdf}

\bibitem{xie2}
G.~Xie. An ID-based key agreement scheme from pairing.
\url{http://eprint.iacr.org/2005/093.pdf}
\end{thebibliography}
\end{document}